\documentclass[aps,
prd,
10pt,
onecolumn,
showpacs,
showkeys,
nofootinbib,
longbibliography,
superscriptaddress,
notitlepage]{revtex4-1}
\usepackage{latexsym} 
\usepackage{amsmath}  
\usepackage{amssymb}
\usepackage{mathtools}
\usepackage{hyperref}
\usepackage{color,graphicx,colortbl}
\usepackage{epsfig}
\usepackage{booktabs}
\usepackage{graphicx}
\usepackage{subfigure} %

\usepackage{xcolor}
\hypersetup{
    colorlinks,
    linkcolor={blue!100!black},
    citecolor={cyan!100!black},
    urlcolor={red!0!black}
}

\newcommand{\sgrA}{Sgr~A$^{*}$ }

\begin{document}

%%%%%%%%%%%%%%%%%%%%%%%%%%%%%%%%%%%%%%%%%%%%%%%%
%%%%%%%%%%%%%%%%%%%%%%%%%%%%%%%%%%%%%%%%%%%%%%%%

\hfill {{\footnotesize USTC-ICTS/PCFT-26-12}}

\title{
The imitation game (r)evolutions: \\ 
$Q$-star effective shadow from GRMHD analysis}

\author{V\'ictor Jaramillo}
\email[]{jaramillo@ustc.edu.cn}
\affiliation{Department of Modern Physics, University of Science and Technology of China, Hefei, Anhui 230026, China}
\affiliation{Departamento de Física, Universidad de Guadalajara, 44430 Guadalajara, Jalisco, México}

\author{Laura Meneses}
\email{laurak@ciencias.unam.mx}
\affiliation{Instituto de Ciencias Nucleares, Universidad Nacional
  Aut\'onoma de M\'exico, Circuito Exterior C.U., A.P. 70-543,
  Coyoac\'an, M\'exico 04510, CdMx, M\'exico}

\author{H\'ector R. Olivares S\'anchez}
\email{h.sanchez@ua.pt}
\affiliation{Departamento de Matemática da Universidade de Aveiro and Centre for Research and Development in Mathematics and Applications (CIDMA), Campus de Santiago, 3810-183 Aveiro, Portugal.}

\author{Carlos Herdeiro}
\email{herdeiro@ua.pt}
\affiliation{Departamento de Matemática da Universidade de Aveiro and Centre for Research and Development in Mathematics and Applications (CIDMA), Campus de Santiago, 3810-183 Aveiro, Portugal.}
\affiliation{Programa de P\'os-Gradua\c{c}\~{a}o em F\'{\i}sica, Universidade 
	Federal do Par\'a, 66075-110, Bel\'em, Par\'a, Brazil.}

\author{Dar\'io N\'u\~nez }%\orcidlink{0000-0003-0295-0053}
\email{nunez@nucleares.unam.mx}
\affiliation{Instituto de Ciencias Nucleares, Universidad Nacional
  Aut\'onoma de M\'exico, Circuito Exterior C.U., A.P. 70-543,
  Coyoac\'an, M\'exico 04510, CdMx, M\'exico}

\author{Shuang-Yong Zhou}
\email[]{zhoushy@ustc.edu.cn}
\affiliation{Interdisciplinary Center for Theoretical Study, University of Science and Technology of China, Hefei, Anhui 230026, China}
%\affiliation{Department of Modern Physics, University of Science and Technology of China, Hefei 230026, China}
\affiliation{Peng Huanwu Center for Fundamental Theory, Hefei, Anhui 230026, China}

%%%%%%%%%%%%%%%%%%%%%%%%%%%%%%%%%%%%%%%%%%%%%%%%
%%%%%%%%%%%%%%%%%%%%%%%%%%%%%%%%%%%%%%%%%%%%%%%%

\begin{abstract}
{\it Q}-stars are a class of boson stars arising in scalar-field theories with 
%(at least) sextic} 
interacting potentials, minimally coupled to gravity. We show that, in certain regions of parameter space, the angular velocity of stable timelike circular geodesics around $Q$-stars can attain a maximum at a nonzero radius. Notably, this behaviour may occur for \textit{stable} configurations. This feature has been argued to produce effective shadows, but so far it has only been investigated for unstable solutions. We test this possibility by performing general relativistic magnetohydrodynamic evolutions for a representative stable $Q$-star model. A low-density, low-luminosity central region is indeed observed to form and persist -- at least until the evolution becomes affected by numerical viscosity. As a proof of principle, this suggests that families of stable bosonic stars can act as black hole mimickers. Moreover, for the model at hand, a heuristic analysis shows that the  effective shadow has a comparable size to that of a  Schwarzschild black hole with the same mass. Importantly, this mechanism for generating an effective shadow does not rely on the object being ultracompact, or an ad hoc chosen accretion disk.
\end{abstract}

\maketitle

\tableofcontents
%%%%%%%%%%%%%%%%%%%%%%%%%%%%%%%%%%%%%%%%%%%%%%%%
%%%%%%%%%%%%%%%%%%%%%%%%%%%%%%%%%%%%%%%%%%%%%%%%

\section{Introduction}

Observations by the Event Horizon Telescope (EHT) have opened a new window into the strong-field regime of gravity, providing horizon-scale images of compact objects and enabling direct tests of the black hole paradigm. The observation of shadow-like features in sources such as M87* and Sgr A* has intensified both theoretical and observational efforts to understand the formation and interpretation of black hole shadows, as well as to assess to what extent alternative compact objects may produce similar observational signatures. In this context, horizonless configurations that can mimic certain properties of black holes have received growing attention.

Following the fruitful developments of perturbative quantum field theory, interest in solitons within relativistic systems gained momentum. These solitons are extended, localized field configurations that are stable. Their stability may arise from topological reasons \cite{Manton:2004tk}, or be ensured by internal symmetries. In the latter case, they are known as non-topological solitons, a class which includes  $Q$-balls \cite{Friedberg:1976me,Coleman:1985ki,Rosen:1968mfz} (see \cite{Zhou:2024mea} for a recent review). To evade Derrick's theorem, $Q$-balls must be time-dependent in field space rather than static. $Q$-balls naturally arise in supersymmetric extensions of the Standard Model \cite{Kusenko:1997zq,Laine:1998rg,Enqvist:2000gq}. In the early universe, they can form as a by-product of Affleck--Dine baryogenesis \cite{Affleck:1984fy,Enqvist:1997si,Enqvist:1998en,Enqvist:1999mv,Kasuya:2000wx,Multamaki:2002hv,Harigaya:2014tla}, and may persist to the present day as dark matter candidates \cite{Kusenko:1997si,Enqvist:1998xd,Banerjee:2000mb,Kusenko:2001vu,Roszkowski:2006kw,Shoemaker:2009kg,Kasuya:2011ix,Kasuya:2012mh,Kawasaki:2019ywz}.

$Q$-balls form due to the attractive nature of the scalar potential. In the simplest case of a single $U(1)$ scalar field, this requires the (effective) potential to include at least a $|\Phi|^6$ term, enabling condensation within field fragments. In the presence of gravitational attraction, non-topological solitons known as boson stars \cite{Kaup:1968zz,Ruffini:1969qy} can form even with a free scalar potential. However, such mini-boson stars are limited in compactness, at least within the stable branch~\cite{Gleiser:1988ih,Seidel:1990jh}. Including self-interactions—leading to what we refer to as $Q$-stars in this paper—enriches the phenomenology by increasing both the mass and the compactness of the boson star, but still allowing stable stars.

Building on the role of self-interactions in boson stars, stationary solutions supported by sextic scalar potentials have been extensively constructed and analyzed in the literature. Static, spherically symmetric configurations were first studied in \cite{Friedberg:1986tq,Lee:1986ts}, while non-spherical solutions, such as dipolar and chain configurations, were obtained in \cite{Herdeiro:2021mol,Gervalle:2022fze}, including their black hole counterparts \cite{Herdeiro:2023mpt}. Rotating solutions were investigated in \cite{Kleihaus:2005me}and also in the case of gauge fields \cite{Jaramillo:2024shi}. The stability properties of these configurations have also been explored in detail \cite{Ildefonso:2023qty,DiGiovanni:2020ror,Siemonsen:2020hcg}, revealing robustness against a variety of perturbations. Interestingly, spinning boson stars, which are unstable in the case of a free scalar potential, can be stabilized once suitable self-interactions—such as those of a $Q$-star–type potential—are included. Recent studies have also shown that these interaction terms enable a novel energy-amplification mechanism for waves scattering off $Q$-balls \cite{Saffin:2022tub,Cardoso:2023dtm,Chang:2024xjp,Zhang:2025oud,Zhang:2025nqr} and $Q$-stars \cite{Cardoso:2023dtm,Gao:2023gof,Chang:2024xjp}, namely a superradiant process driven by internal-space rotation. Additionally, $Q$-balls and $Q$-stars can exhibit rich spatial configurations that go beyond spherical symmetry \cite{Copeland:2014qra,Xie:2021glp,Xie:2023psz,Jaramillo:2024smx,Jaramillo:2024cus}.

The possibility that boson stars may produce strong gravitational lensing has motivated early studies of their lensing properties and possible shadow-like features~\cite{Cunha:2015yba,Cunha:2016bjh,Cunha:2017wao,Herdeiro:2021lwl}. The latter raised the question  whether their observational signatures may resemble those of black holes; this led to pioneering studies of such \textit{imitation game}~\cite{Olivares:2018abq,Herdeiro:2021lwl}, reloaded~\cite{Sengo:2024pwk} and re-analysed~\cite{,Rosa:2022tfv,Rosa:2023qcv} in different setups to identify the conditions under which boson stars can generate central dim regions or effective shadows, even in the absence of an event horizon. These analyses highlight both similarities and qualitative differences with the standard black hole scenario, and emphasize the role played by the spacetime structure and geodesic motion, as well as the astrophysical light source, in shaping observable features. But general-relativistic magnetohydrodynamic evolutions (GRMHD) simulations, necessary to produce state of the art synthetic images in this context, were only studied in~\cite{Olivares:2018abq}, where the boson star models that could mimic black hole images were unstable. Thus, it remains to see if stable boson stars could actually be players in this imitation game, under GRMHD evolutions. An affirmative answer to this question will be provided in this paper.

More broadly, boson stars belong to a larger class of horizonless or exotic compact objects that may act as black hole mimickers. These include, among others, black holes with scalar~\cite{Herdeiro:2014goa} or vector hair~\cite{Herdeiro:2016tmi}, Proca stars~\cite{Brito:2015pxa}, and other self-gravitating solitonic configurations. Assessing the observational degeneracies and discriminating signatures among these possibilities is a central goal of current strong-gravity astrophysics.

In this work we investigate a specific mechanism through which $Q$-stars can produce effective shadow-like features without requiring ultracompactness ($i.e.$ possessing light rings, which may source a spacetime instability~\cite{Keir:2014oka,Cardoso:2014sna,Cunha:2022gde} - see also~\cite{Benomio:2024lev,Marks:2025jpt,Evstafyeva:2025mvx,Cunha:2025oeu} for the ongoing debate). We focus on configurations for which the angular velocity of stable timelike circular geodesics peaks at a nonzero radius and explore the observational consequences through GRMHD evolutions. Our results indicate the formation of a persistent low-density, low-luminosity central region, supporting the interpretation of $Q$-stars as viable black hole mimickers in certain regimes.

The remainder of this paper is organized as follows. In Sec.~\ref{sec:spacetime} we present the theoretical framework and the $Q$-star models under consideration and we describe the geodesic structure and the conditions leading to a non-monotonic angular velocity profile. In Sec.~\ref{sec:GRMHD} we outline the GRMHD setup and numerical methodology and present our results and their observational implications. We conclude in Sec.~\ref{sec_discussion} with a summary and outlook.

Throughout this work we adopt the metric signature $(-,+,+,+)$ and use units in which $c=1$. Newton's constant is set to $G=1$, except in selected expressions where it is retained for clarity in the presentation of the physical model.

%%%%%%%%%%%%%%%%%%%%%%%%%%%%%%%%%%%%%%%%%%%%%%%%
%%%%%%%%%%%%%%%%%%%%%%%%%%%%%%%%%%%%%%%%%%%%%%%%

\section{\texorpdfstring{$Q$}{Q}-star spacetime}
\label{sec:spacetime}

The background spacetime for the GRMHD simulations will be that of a self-gravitating $Q$-ball, also known as a $Q$-star. These configurations, composed of a complex scalar field, can be interpreted as boson stars but with a different scalar potential compared to the simpler mini-boson stars \cite{Kaup:1968zz,Ruffini:1969qy}. They are also referred to in the literature as soliton stars \cite{Friedberg:1986tq,Lee:1986ts}.
In flat spacetime, one of the simplest potentials that can support $Q$-balls\footnote{This applies to the case of non-topological solitons in 3+1 dimensions within an effective field theory involving a single scalar field \cite{Coleman:1985ki}. A model without higher dimensional effective operators can be easily written down when allowing for two scalar fields, as considered in the earlier work \cite{Friedberg:1976me}.} is given by a complex scalar field with the potential
$
    V(\Phi) = \mu^2|\Phi|^2 + \lambda/2|\Phi|^4+ \nu/3|\Phi|^6 \, .
$
This is, in fact, the simplest polynomial potential with symmetry $\mathbb{Z}_2$ that supports $Q$-ball solutions when $\lambda$ is negative. In the absence of gravitational attraction, the (nonlinear) interaction terms of the potential must drip below zero in order for $Q$-balls to form. Negative interaction terms effectively induce attractive forces among scalar perturbations and energetically favor localized field configurations over homogeneous ones. This is because localization allows the field to attain large amplitudes in its core, where the negative interaction terms contribute more significantly and reduce the total energy. Physically, this corresponds to particles clustering together to form bound, localized structures.

$Q$-stars are obtained by minimally coupling a complex scalar field to gravity, where the scalar field acts as the source of the gravitational field through the energy-momentum tensor:
\begin{equation}\label{eq:energy_momentum_tensor}
 T_{\mu\nu} = 2\partial_{(\mu}\Phi^\dagger\partial_{\nu)}\Phi - g_{\mu\nu} \left(\partial^\kappa\Phi^\dagger\partial_\kappa\Phi+V(|\Phi|)\right) \, ,
\end{equation}
where $^\dagger$ denotes complex conjugation, and $g_{\mu\nu}$ is the spacetime metric, determined by the Einstein equations. These gravitational solitons exhibit several interesting properties. Firstly, their masses and compactness can reach significantly higher values compared to their simpler mini-boson star cousins, arising from a potential consisting only of a mass term,  $V = \mu^2 |\Phi|^2$. Moreover, even when compared to quartic self-interacting boson stars \cite{Colpi:1986ye}, these $Q$-stars can be more compact (see \cite{Amaro-Seoane:2010pks} for a direct comparison). Their radius can be as small as their associated Schwarzschild light ring, as demonstrated in \cite{Kesden:2004qx,Macedo:2013jja}, or can even approach the Schwarzschild radius when using an alternative definition of the radius \cite{Kleihaus:2011sx}.

Secondly, $Q$-stars have appealing stability properties. For example, the sequence of solutions can form two disconnected stable branches in both spherical \cite{Kleihaus:2011sx,Kusmartsev:1990cr} and dipolar configurations \cite{Ildefonso:2023qty}. Additionally, for rotating boson stars, certain stable regions—though not full branches—have been identified \cite{Siemonsen:2020hcg,Chang:2024xjp}, whereas mini-boson stars are unstable in the rotating case \cite{DiGiovanni:2020ror}. Finally, these boson star spacetimes have been shown to release energy and angular momentum when interacting with incident waves, meaning that they can participate in superradiant processes \cite{Saffin:2022tub,Gao:2023gof,Chang:2024xjp}. $Q$-stars are a class of dynamically robust compact objects capable of mimicking the observational appearance of a black hole. In particular, as discussed in the following sections, they can produce an effective shadow comparable in size to that of a black hole.

In \cite{Olivares:2018abq}, the observational appearance of a non-rotating mini-boson star was studied using GRMHD simulations followed by radiative transfer calculations. Two different boson star spacetimes were considered, with a key distinction in their background properties that ultimately influenced the simulations. Specifically, the angular velocity profile $\Omega$ for circular (timelike) geodesics exhibited an inner region where $\Omega$  decreased toward the center in one configuration, while no such region existed in the other. The presence of this inner region, where magnetorotational instability is suppressed (as will be discussed in the next section), was found to play a crucial role. In particular, the authors noticed that the location of the maximum of $\Omega$ coincided with the inner radius of a small torus of stalled plasma, leading to the formation of a dark region at the center of one of the horizonless and surfaceless object explored while for the other one, with monotonically decreasing profile of $\Omega$, the plasma accumulated at the center.

We build upon the hypothesis proposed in \cite{Olivares:2018abq} and conjecture that the existence of dark central regions surrounded by bright emission in black hole mimicker spacetimes depends on the presence of an inner region with $\Omega$ increasing as a function of the radius $r$. Moreover, we expect that the farther this maximum is from the center of the mimicker, the larger the resulting shadow will be. For mini-boson stars, only certain configurations within the unstable branch were found to exhibit an angular velocity profile $\Omega$ (cf.~Eq.~\eqref{eq:Omega}) with a local maximum for $r > 0$. This raises the question of whether it is possible to obtain a stable configuration in a model that could also produce an expected shadow. To complete this two-fold task, we now explore $Q$-stars as a simple model capable of realizing this scenario. We find that a quartic term ($\lambda$) in the potential alone is insufficient; however, the interplay between the quartic and sextic terms allows for the construction of a successful configuration.

\subsection{Families of $Q$-stars and the role of self-interactions}

We begin by defining some convenient rescalings of the parameters $\lambda$ and $\nu$ which can be derived from the rescaling of the full system by considering $1/\mu$ as a length and time unit. In particular, we use the same conventions as in \cite{Chang:2024xjp}:
\begin{equation}
    g=\frac{\lambda M_{\rm P}^2}{\mu^2} \, , ~~ h=\frac{\nu M_{\rm P}^4}{\mu^2} \, , 
\end{equation}
defining new coordinates $\tilde{x}^a=x^a \mu$, the field $\tilde{\Phi}=\Phi/M_{\rm P}$, where $M_{\rm P}=1/\sqrt{8\pi G}$ is the reduced Planck mass. After rescaling and removing the tilde symbols from $\tilde{x}^a$ and $\tilde{\Phi}$ for simplicity, the equations of motion that determine $\Phi$ and $g_{\mu\nu}$ conform the Einstein-Klein-Gordon system and take the (dimensionless) form ${G^\mu}_\nu=8\pi G{T^\mu}_\nu$ and $(\nabla^\kappa\nabla_\kappa-V')\Phi=0$, where the energy momentum is given by Eq.~\eqref{eq:energy_momentum_tensor}, the scalar potential by
\begin{equation}
V(|\Phi|) = |\Phi|^2 + \frac{1}{2}g|\Phi|^4 + \frac{1}{3} h|\Phi|^6 \, ,
\end{equation}
and we have defined $V':=dV/d|\Phi|^2 = 1+ g|\Phi|^2 + h|\Phi|^4$. The model has two independent parameters, $g$ and $h$. With this in mind, we proceed to solve the Einstein-Klein-Gordon equations to find stationary solutions. To construct spherical static boson stars, we consider:
\begin{equation}\label{eq:ansatz}
    \Phi=\phi(r) e^{-i\omega t} \, , ~~ ds^2 = -e^{2F_0(r)}dt^2 + e^{2F_1(r)}\left(dr^2+r^2d\Omega^2\right) \, .
\end{equation}
with $\omega$ a real number to be determined by the boundary conditions and corresponding to the frequency of pulsation of the scalar field. From the metric ansatz in \eqref{eq:ansatz}, we see that the background is solved in the isotropic gauge.

We solve numerically for the functions $\phi$, $F_0$, $F_1$ and the unknown $\omega$ using the equations fully displayed in \cite{Chang:2024xjp} and using the spectral code described in \cite{Alcubierre:2021psa}, splitting the $r$ domain in 8 shells and using between 18 and 24 spectral coefficients. For each set of values $(g,h)$ we construct sequences of solutions by varying the value\footnote{
Depending on the values $(g,h)$, sometimes $F_0(0)$ is not a good parameter to parametrize the sequence of solutions. For these cases we vary $\omega$ instead. 
} $F_0(0)$ starting from values $F_0(0)\to0^-$, corresponding to the Newtonian limit. We calculate the value of the total mass $M$ of the configurations by extracting it from the asymptotical behavior of the metric functions in Eq.~\eqref{eq:ansatz}. The radius $R$, often used to describe the size of a boson star, is defined as the areal radius enclosing 99\% of the total mass. More precisely, in the coordinates given by \eqref{eq:ansatz}, this radius is expressed as $R = e^{F_1(r_{99})} r_{99}$, where $r_{99}$ is the (isotropic) radial coordinate $r$ at which the Misner-Sharp mass function, $M(r) = -2r/(1 + r \partial_r \ln \Psi)$, satisfies $M(r_{99}) = 0.99M$. Both quantities, the mass and radius of some families of solutions of $Q$-stars are displayed in Fig.~\ref{fig:M-omega}. In this figure and from this point onward, we will use units such that $G = 1$.

\subsection{Selection of the configurations}

\begin{figure}[ht!]
\includegraphics[width=0.49\textwidth]{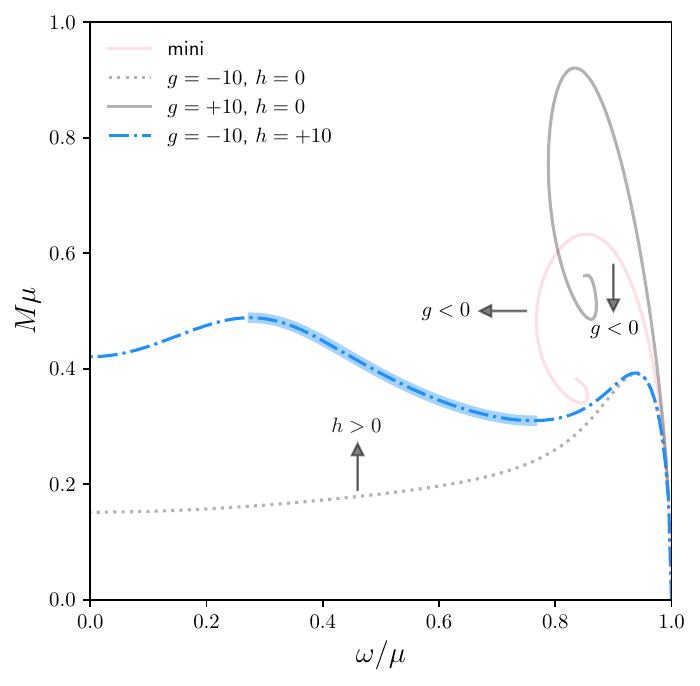}\includegraphics[width=0.482\textwidth]{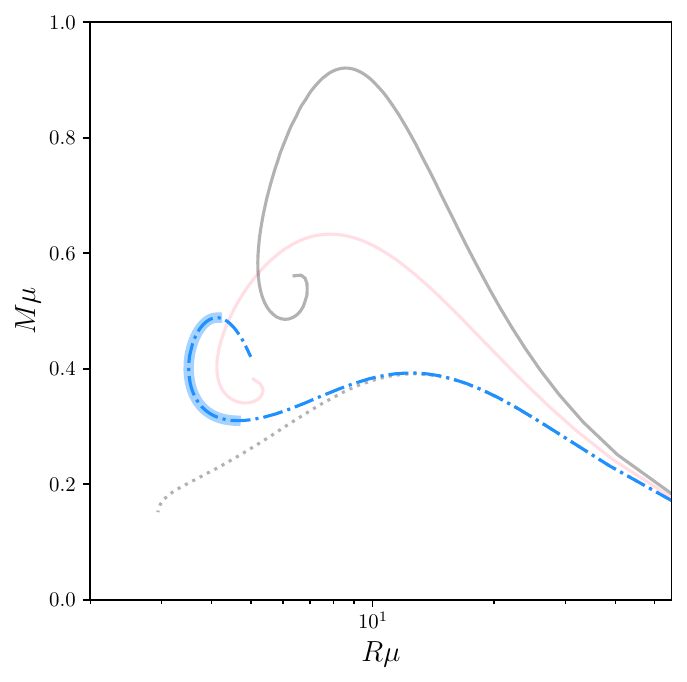} 
 \caption{$Q$-star sequences for different values of the self-interacting parameters $g$ and $h$. We include for comparison the mini-boson star and show with arrows the effect of including a negative $g$ (for positive $g$ the behavior follows the opposite direction). Also with an arrow we indicate the effect of including a positive $h$ in the strong field region in the sequences of negative $g$. The thick blue curve denotes the second stable region: the relativistic stable branch.}
    \label{fig:M-omega}
\end{figure}

To better understand the role of the parameters $g$ and $h$ in shaping the properties of the resulting boson stars, we begin by considering the Newtonian limit, where $\Phi \to 0$. This corresponds to the limits $M \to 0$, $\omega \to \mu$, and $R \to \infty$ in Fig.~\ref{fig:M-omega}. In this regime, the influence of the higher-order interaction terms, parameterized by $g $ and $h$, is negligible, and all families of solutions exhibit similar behavior as they approach this limit.

Next, we compare the mini-boson star case ($g=0=h$) with configurations that have either positive or negative values of $g$ keeping the sextic term zero. When $g$ is positive, corresponding to a repulsive self-interaction, the stars become more massive as $g$ increases, with their mass scaling as a power of $g$ \cite{Colpi:1986ye}. However, their compactness and frequency profiles approach an asymptotic limit \cite{Colpi:1986ye}. Notably, for positive $g$, the spiral structure in the $M$ vs. $\omega$ diagram does not unroll indefinitely as $g$ grows. Instead, all configurations remain confined within the range $0.8\mu \lesssim \omega < \mu$, as illustrated by the $g=+10$ gray curve in Fig.~\ref{fig:M-omega}. An important aspect of these $M$ vs.  $\omega$  diagrams is that stable boson stars lie along stable branches, defined by regions where $dM/d\omega < 0$, extending from the Newtonian limit at $M \to 0$ to the global minimum of $\omega$ - see~\cite{Santos:2024vdm} for a generic discussion of perturbative stability of spherical bosonic stars. Quartic self-interacting boson stars possess only a single stable branch, as can be confirmed in the figure.

For negative $g$, the self-interaction becomes attractive, leading to less massive configurations \cite{Barranco:2010ib} and causing the spiral to unroll. If $|g|$ is sufficiently big, such as the $g=-10$ gray curve in Fig.~\ref{fig:M-omega}, the frequency can extend down to $\omega \to 0$. This behavior is represented using arrows in the left panel of Fig.~\ref{fig:M-omega}. In addition to these effects, a negative $g$ shifts the maximum mass toward higher values of $\omega$. When extending the solutions into the strong-field regime (large $|\Phi|$), introducing a positive value of $h$ provokes additional repulsive interaction that becomes significant at larger $|\Phi|$, creating conditions for a second region of stability. This second stable region, referred to as the \textit{relativistic stable branch}, is illustrated in Fig.~\ref{fig:M-omega} for the case $g=-10$ and $h=+10$ (thick blue curve). We have verified the stability of several configurations along this curve using full 3+1 numerical relativity, particularly for cases with $\omega/\mu = \{0.3,0.4,0.5,0.6,0.7\}$, employing the \texttt{Einstein Toolkit} \cite{EinsteinToolkit:web} as implemented in \cite{Chang:2024xjp}.

As noted earlier, mini-boson stars exhibit a local maximum of the angular velocity $\Omega$ at $r > 0$ only in the unstable branch, meaning that configurations most relevant for mimicking a black hole reside deep within the ``spiral'' region of solutions where the location of the maximum is %farther apart 
far away from the center. For the family of  $Q$-stars with negative $g$ and positive $h$, however, as mentioned above, the spiral unrolls, creating the relativistic stable branch. Furthermore, as shown in the right panel of Fig.~\ref{fig:M-omega}, the masses in this relativistic stable branch remain comparable to those of other boson stars, but their radii $R$ are significantly smaller, making these configurations more \textit{compact}. This leads us to the question of effective shadow formation: are these stable configurations also likely to possess local maxima of $\Omega$? %\vjp{I forgot to mention that the weak energy condition is satisfied in the relativistic stable branch.}

\begin{figure}  
\subfigure[~~]{
\includegraphics[width=0.4\textwidth]{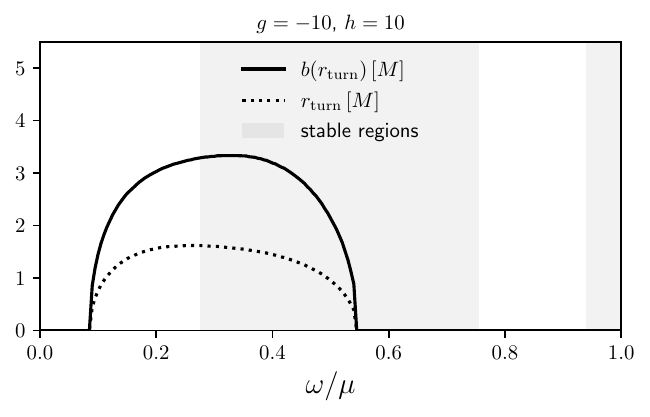}~~\includegraphics[width=0.4\textwidth]{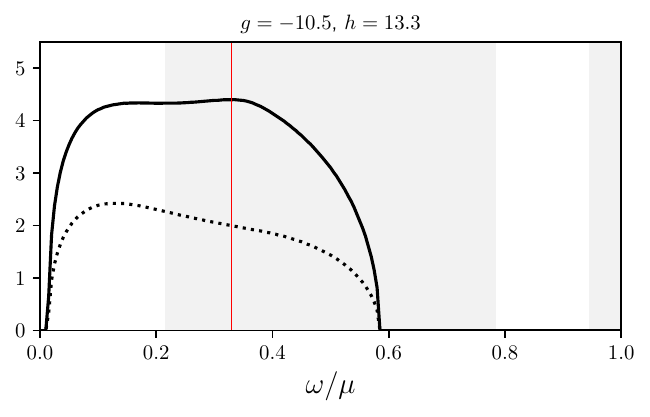}
}
\subfigure[~~]{
\hspace{-0.9cm}\includegraphics[width=0.435\textwidth]{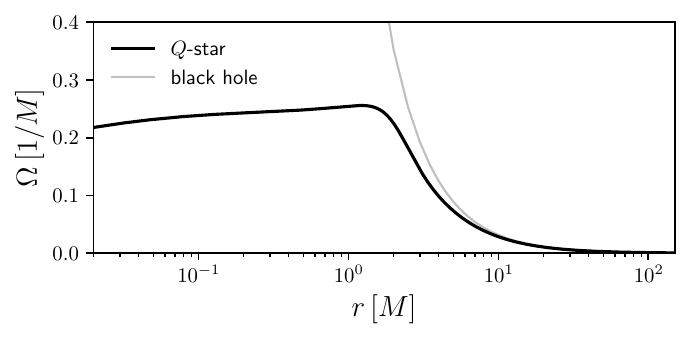}~~\includegraphics[width=0.4\textwidth]{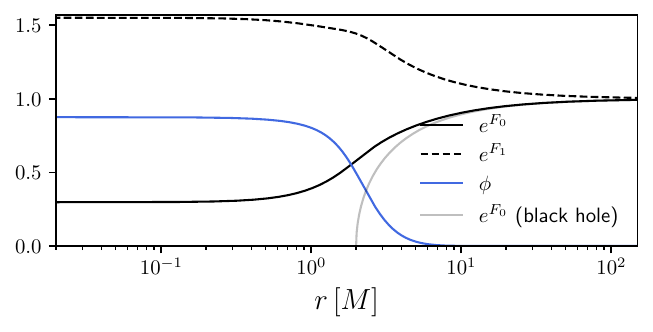}
}
\caption{Selection of the reference $Q$-star. Panel (a) shows the radius and impact parameter corresponding to the location of the local maximum in the angular frequency profile for two sequences of $Q$-stars. Notably, two stable regions emerge, corresponding to the stable branches of the background metric identified in Fig.~\ref{fig:M-omega}. A thin red line marks the reference $Q$-star used in the GRMHD simulations. Panel (b) presents the angular velocity, metric functions, and scalar field profile for this reference  Q -star, alongside comparisons with a Schwarzschild black hole. In this figure, $r[M]$ is not the radial coordinate \eqref{eq:ansatz}, but the areal radius in units of $M$, \textit{i.e.}, $r[M] := e^{F_1}r/M$.}
    \label{fig:Omega}
\end{figure}
The magnitude of the angular velocity, as measured at infinity, of a timelike circular orbit in the spacetime of a spherical $Q$-star represented by the metric \eqref{eq:ansatz} is\footnote{
See, \textit{e.g.}, Eq.~(8) of \cite{Sengo:2024pwk} applied to the metric \eqref{eq:ansatz}.
}
\begin{equation}\label{eq:Omega}
    \Omega(r) := \left|\frac{d\varphi}{dt}\right|= e^{F_0-F_1}\left(\frac{F_0'}{r+r^2F_1'}\right)^{1/2} \, .
\end{equation}
Local maxima at $r>0$ can be identified by evaluating the angular velocity profiles, $\Omega$, for all solutions obtained for each set of parameters $(g,h)$. We denote the areal radius at which $\Omega$ reaches this turning point as $r_{\rm turn}$. In the region inside $r_{\rm turn}$, \textit{i.e.} where $d\Omega/dr > 0$, the magnetorotational instability is suppressed \cite{Balbus:1991ay}. In this way, using the metric coefficients, it is possible to estimate the presence of a region of luminosity depletion, even if the non-gravitational part of the central body does not interact with light, as is the case of a scalar star.

We first reproduce the known results for the location of this turning point in the case of a mini-boson star \cite{Olivares:2018abq}. Our focus is then turned to the sequence of solutions with $g=-10$ and $h=+10$, confirming that a turning point indeed exists. Moreover, we find that a significant portion of this region (in this case, more than half) intersects with the relativistic stable branch -- see the left panel of Fig.~\ref{fig:Omega} (a). Note that in this plot, we choose the total mass of the star, $M$, instead of $\mu$, as the units for the dimensionless quantities such as $r_{\rm turn}$.

The values of $r_{\rm turn}$ for this $Q$-star sequence with $g=-10$ and $h=+10$ are comparable to those found in unstable mini-boson stars, where a torus forms without penetrating the area with $d\Omega/dr>0$. In such cases, ray-traced images revealed a dark region, although relatively smaller than that of a Kerr black hole \cite{Olivares:2018abq}. An estimation of the expected size of such dark region is given by the impact parameter for a light ray grazing at $r_{\rm turn}$, $b(r_{\rm turn})= r_{\rm turn}e^{-F_0(r_{\rm turn})+F_1(r_{\rm turn})}$. We find that both $r_{\rm turn}$ and the corresponding impact parameter $b(r_{\rm turn})$  can be increased by exploring some close values of $(g,h)$. After a brief survey of a dozen solution families, we identify the sequence with $g = -10.5$, $h = 13.3$. In this case, the solution with $\omega = 0.33\mu$ attains a notably high impact parameter, $b(r_{\rm turn}) = 4.4$ (see the right panel of Fig.~\ref{fig:Omega}). We designate this solution as our reference $Q$-star, characterized by $r_{\rm turn} = 2.0$ with  $\Omega = 0.081$, a total mass of $M = 0.624$, and compactness $M/R = 0.164$. The profiles of $\Omega$, the metric functions, and the scalar field are shown in panel (b) of Fig.~\ref{fig:Omega}, along with a comparison to the corresponding Schwarzschild solution of identical mass. Also for comparison, note that the impact parameter of the light ring of the Schwarzschild black hole is $b=\sqrt{27}M\simeq 5.2M$.
\begin{figure}
%[H]    
\includegraphics[width=0.7\textwidth]{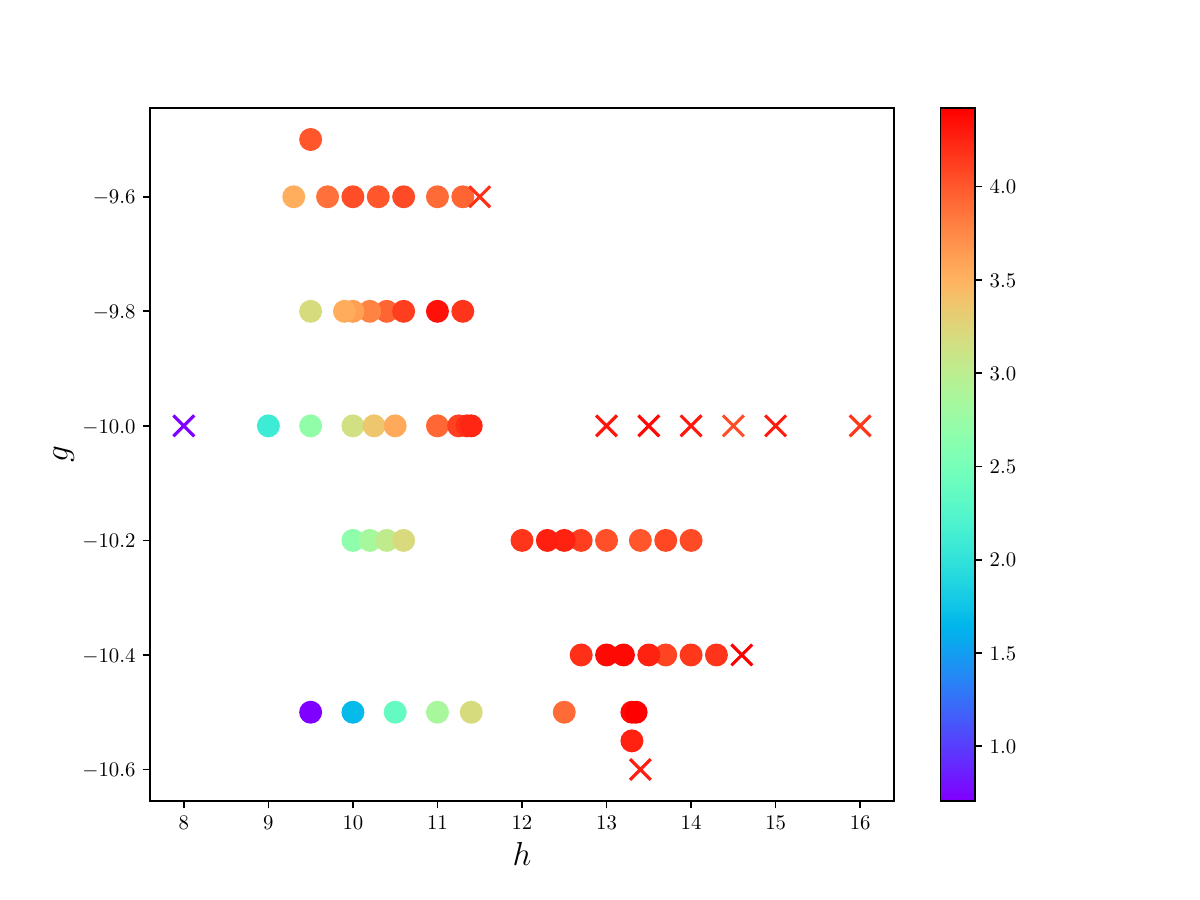}
 \caption{Map plot for values of $g$ and $h$ with a relativistic stable region around $(g=-10,h=12)$. Different colors are used to denote the largest values of the impact parameter $b(r_{\rm turn})$
 %cross section 
 for the corresponding values of $g$ and $h$. Crosses indicate that such values are not in the stable branch.}
    \label{fig:g-h}
\end{figure}

We have seen that the configuration with $\left(g=-10, h=10\right)$ has a relativistic stable branch. In Fig.~\ref{fig:g-h}, we explore the values of $g$ and $h$ around that point. For each pair of values, we identity the largest possible impact parameter by constructing spacetime configurations for several values of $\omega$. We use a color code to indicate such largest value for each pair in Fig.~\ref{fig:g-h}. Notice that the plot seems to indicate a zone of $\left(g,h\right)$ points with large impact parameters.

As mentioned, in presenting the results in Fig.~\ref{fig:Omega} and Fig.~\ref{fig:g-h}, we have used $M$ instead of $\mu$ as units to construct dimensionless quantities. This choice is particularly convenient when treating the $Q$-star as a black hole mimicker, as will be explored in the next section.
To facilitate numerical simulations, we have prepared tabulated data for the background solution of our reference $Q$-star and interpolated it into the areal gauge, given by $ds^2 = -e^{2F_0} dt^2 + a^2 dr_a^2 + r_a^2 d\Omega^2$. These interpolated fields are then fed to the GRMHD code as functions of the dimensionless radius\footnote{
    In the code we deal with the dimensionless radius $\mu r$ and receive as one of the outputs the dimensionless mass $\mu M$. Therefore the dimensionless radius can be expressed as $r_a/M=e^{F_1}(\mu r)/(\mu M)$.
} $r_a/M$. In the following section we will drop the $a$ subindex and use $r$ to represent the areal radius.

%%%%%%%%%%%%%%%%%%%%%%%%%%%
%%%%%%%%%%%%%%%%
\section{General relativistic magnetohydrodynamic simulations}
\label{sec:GRMHD}

\subsection{Simulation setup}
\label{sec:simulation-setup}
As our spacetime background, we adopt the reference $Q$-star identified in the previous section, corresponding to the solution with $\omega = 0.33\mu$ and couplings $(g,h)=(-10.5,\,13.3)$. This configuration lies on the relativistic stable branch and is distinguished by the large radius at which the angular velocity reaches its maximum, yielding a correspondingly large impact parameter.
We perform a 3D simulation of magnetized accretion onto this reference $Q$-star.
The simulation
evolves the equations
of GRMHD
using the code \texttt{BHAC}
(\url{https://bhac.science}, see \cite{Porth2016,Olivares2019a} for more details on
the code and the
formulation of the GRMHD equations used).
As initial conditions, we use a torus with constant angular momentum $\ell=-u_\phi/u_t$ \citep{Abramowicz1978}
such that its inner radius is at $r=18\ M$ and its density maximum is at $r=25\ M$,
where $r$ is the areal coordinate.
We choose such a location
for the torus
%(perhaps unusually distant for SANE simulations)
%\ho{Search for a reference with similarly large distance}
to ensure that material arriving
in the vicinity of the $Q$-star smoothly falls towards smaller radii
due to the magneto-rotational instability (MRI) and no big transients are present.
To trigger turbulence through the MRI,
we set up a poloidal magnetic field loop described by the vector potential
\begin{equation}
	A_\phi \propto \max(0,\rho/\rho_{\rm max} - 0.2) \,,
\end{equation}
where $\rho$ is the particle number density in the fluid frame
and $\rho_{\rm max}$ is its maximum value in the initial torus,
and perturb the solution by adding white noise with an amplitude of 4\% to the pressure.
The initial magnetic field is such that the ratio between the highest thermal pressure 
and the highest magnetic pressure in the torus is $p_{\rm max}/p_{\rm mag,max} = 1000$.
The fluid obeys an ideal gas equation of state with adiabatic index $\hat{\gamma} = 4/3$.
To emulate vacuum regions
outside the torus,
the rest of the domain
is filled with
a tenuous atmosphere
following the profiles
$\rho_{\rm atmo} = \rho_{\rm min}(r + r_0)^{-3/2}$
and 
$p_{\rm atmo} = p_{\rm min}(r + r_0)^{-5/2}$,
where $r_0= 1\ M$,
$\rho_{\rm min} =
10^{-5}\rho_{\rm max}$
and 
$p_{\rm min} =
10^{-7}p_{\rm max}$.
These profiles are also
used to reset $p$ and
$\rho$ whenever they fall
below the atmosphere values
during the course of the simulation.
We also
evolve a passive tracer
$\rho_{\rm tr}$
that is advected with
density.
This tracer is initialized
at $\rho_{\rm tr} =1$ inside
the torus and
at $\rho_{\rm tr} =0$ outside of it. We reset
$p$ and $\rho$ to their
atmosphere values
whenever
$\rho_{\rm tr} < 0.1$
in order to prevent
artificial accumulation of
atmosphere material
at the origin.

%{\it Numerical methods.}
We employ a finite volume (FV)  scheme
with a total variation diminishing Lax-Friedrichs (TVDLF)
Riemann solver (also known as Rusanov fluxes)
with piecewise parabolic reconstruction (PPM) \citep{colella_piecewise_1984} and a 
simple two-step predictor-corrector time integration.
To evolve magnetic fields while keeping the solenoidal constraint fulfilled,
we employ the upwind constrained transport scheme of \cite{zanna_echo_2007}.

The simulation is performed in spherical coordinates, with the domain
extending over $r/M\in[0,2500]$. Internally,
we employ a logarithmic radial coordinate $s$,
related to the areal coordinate $r$ by the transformation $r=\exp{(s)} - 1$.
In order to ensure sufficient resolution of the MRI while avoiding
small time steps near the origin due to the Courant-Friedrichs-Lewis (CFL) condition,
we employ four levels of fixed mesh refinement.
Cells in direct contact with the origin are kept at base resolution
($N_r\times N_\theta \times N_\phi = 64\times 16 \times 16$),
while parts of the domain at the highest refinement level
are evolved at effective resolution of $512\times 128 \times 128$.

\begin{figure}
	\centering
    \includegraphics[width=0.8\linewidth]{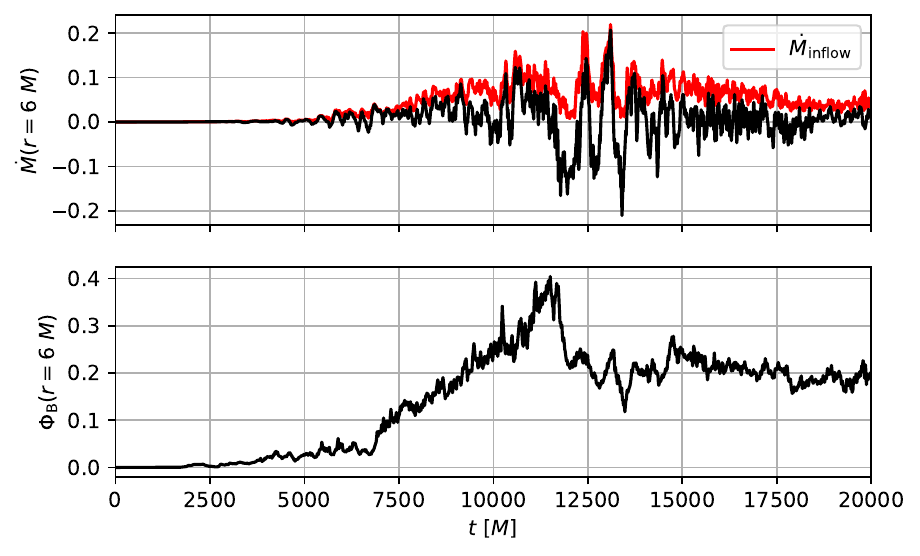}
	\caption{Time series of the mass accretion rate $\dot{M}$
		and radial magnetic flux $\Phi_{\rm B}$ through a
		spherical surface at $r=6\ M$.}
	\label{fig:radial_fluxes}
\end{figure}

\begin{figure}
	\centering
	\includegraphics[width=0.8\linewidth]{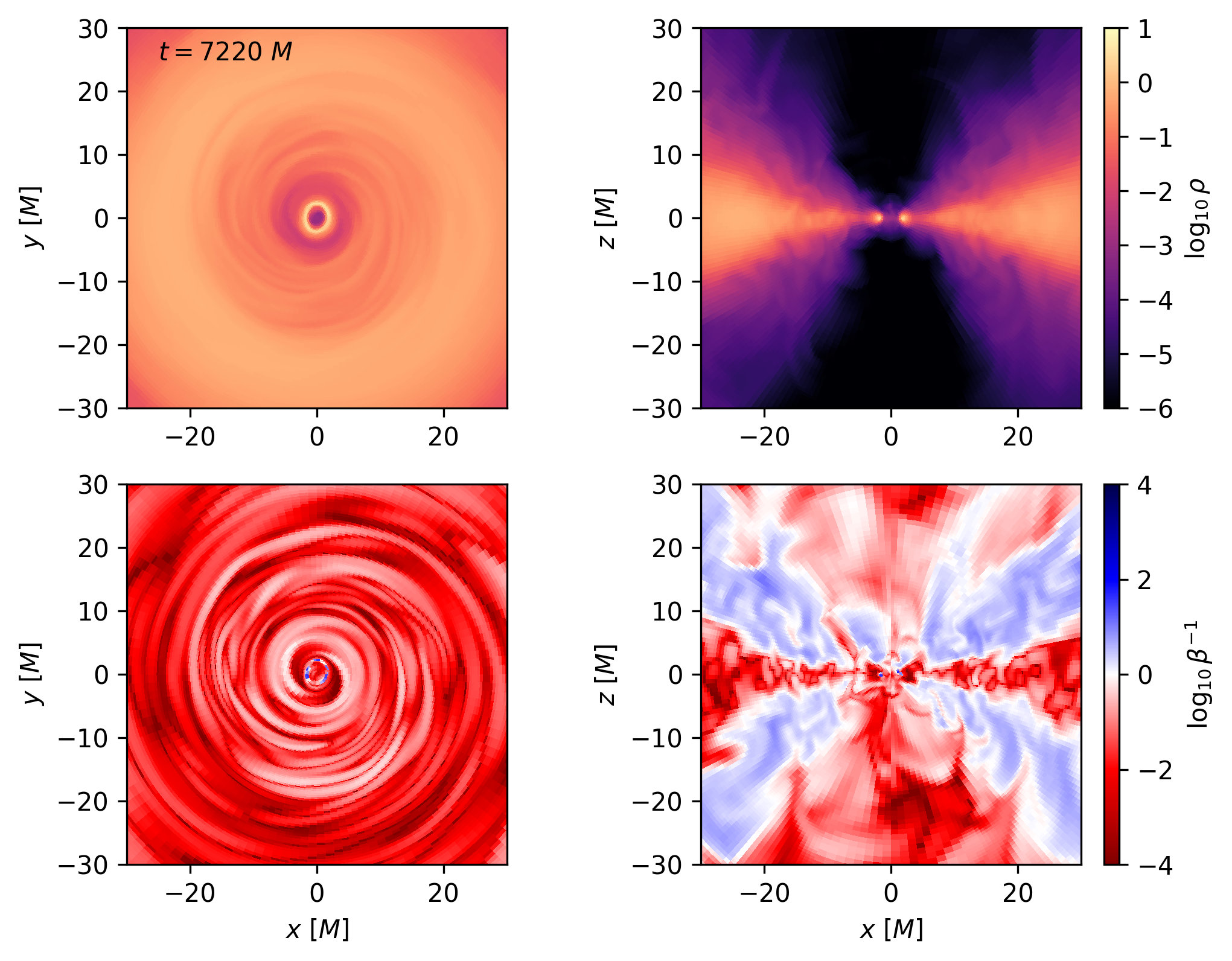}
	\caption{Snapshot of the simulation at $t=7220\ M$, showing density in code units ({\it top panels}) and plasma beta magnetization $\beta \coloneqq p_{\rm gas}/p_{\rm mag}$({\it bottom panels})
	on the equatorial plane ({\it left panels}) and the meridional plane ({\it right panels}).}
	\label{fig:snapshot_0722}
\end{figure}

\subsection{Results}
\label{sec:simulation-results}

\subsubsection{Simulation results}
During the initial
part of the simulation,
the behavior shown is
similar to that observed
for black holes 
in the `Standard And Normal
Evolution' (SANE) regime:
turbulence develops
as a result of the MRI
and material slowly falls
into progressively smaller
orbits \cite{porth_event_2019}.
However, different
from the black hole case,
instead of disappearing
behind the event horizon,
matter accumulates
close to the center of the
boson star.
This accumulation
initially
does not occur at the
center, but slightly
away from it,
forming a torus
around the location predicted
in the previous Sections. This is also
similar to what 
was observed in
\cite{Olivares:2018abq}
for mini-boson stars.
Eventually, material
diffuses to the empty
space inside the ring,
consistent with the
numerical viscosity
estimated for the grid
resolution used.
In the rest of this Section
we provide a few diagnostics
for a more quantitative
description of the
simulation behavior.

We quantify the mass
accretion rate $\dot{M}$
and the absolute value of
the magnetic flux $\Phi_{\rm B}$
through a spherical surface,
defined as

\begin{align}
\label{eq:Mdot}
\dot{M}(r) &\coloneqq -\int_{0}^{2 \pi} \int_{0}^{\pi} \rho u^r \sqrt{-g} \ d\theta \ d\phi \,,\\
\label{eq:PhiB}
\Phi_{\rm B}(r)  &\coloneqq \frac{1}{2}\int_{0}^{2 \pi} \int_{0}^{\pi} |^* F^{rt}| \sqrt{-g} \ d\theta \ d\phi \,,
\end{align}
where $u^r$ is the radial
component of the fluid 4-velocity, $^{*}F^{\mu\nu}$
is the dual of Faraday's
tensor, and $g$ is the
metric determinant.

%We choose to evaluate
%these quantities at
%$r=6\ M$, where
%the metric is still
%approximately similar to
%that of a Schwarzschild BH.

Fig. \ref{fig:radial_fluxes} shows time series of these quantities through a spherical surface at $r=6\ M$.
Matter starts accreting through that radius around $t/M = 2500$; however,
while for BH simulations
$\dot{M}$ remains always positive, in our simulation
it displays large oscillations around a small positive value.
This behavior,
similar to that
observed in
\cite{Olivares:2018abq},
indicates that only a small
fraction of accreting matter
actually accumulates inside
in the region $r < 6\ M$,
while the rest of it
escapes in the form of
a barionic wind.
To show this, we also plot
in Fig.~\ref{fig:radial_fluxes}
the total mass
inflow, calculated as
\begin{equation}
    \label{eq:Mdot-inflow}
\dot{M}_{\rm inflow}(r) \coloneqq -\int_{0}^{2 \pi} \int_{0}^{\pi} \rho \min( u^r,0) \sqrt{-g} \ d\theta \ d\phi \,,
\end{equation}
from which it can be seen
that there is a
steady inflow of matter
even during the times in
which the total mass
accretion rate
becomes negative.
%\ho{Mass accretion rate looks too small. Accretion seems balanced by outflow.
%	Would a barionic outflow produce an observable signature?}

The magnetic flux
displays an initial increase,
after which it
becomes quasi-stationary around $t/M \sim 12000$.
The usual definition
of the `MAD parameter'
\cite{tchekhovskoy_efficient_2011}
$\phi_{\rm MAD} \coloneqq
\Phi_{\rm B}/\sqrt{\langle\dot{M}\rangle}$
at this quasi-stationary
phase would classify
the system within the
MAD regime, $\phi_{\rm MAD} \gtrsim 15$ (in Lorentz-Heaviside units, which we use in this work); however,
the simulation does not
show the magnetic
flux eruptions that
characterize the
magnetically arrested disk
(MAD) state.
This indicates that
the usual expression
for the MAD parameter
may not be adequate to
classify systems
with important outflows
that drive the
mass accretion rate close
to zero, as may
happen in horizonless
exotic compact objects.

We now analyze
the morphology
of the accretion flow
using representative
2D slices.
From $t\sim 7000$, density maps show the formation of a stalled torus
near the predicted location of $r_{\rm turn}=2\ M$.
This torus shows noticeable oscillations due to epicyclic motion
that distorts it into an elliptic shape (Fig. \ref{fig:snapshot_0722}).
However, it appears remarkably stable and survives with a significant
density contrast
for several thousands of $M$ in time,
until matter starts to diffuse to the central hollow region.

% beyond $t\sim 15000\ M$ ,
% when matter starts diffusing into the formerly hollow region.

%{\it Quantitative: size of the 'hollow'.}
%Hollow smaller than predicted: pressure gradient.

\begin{figure}
	\centering    \includegraphics[width=0.9\linewidth]{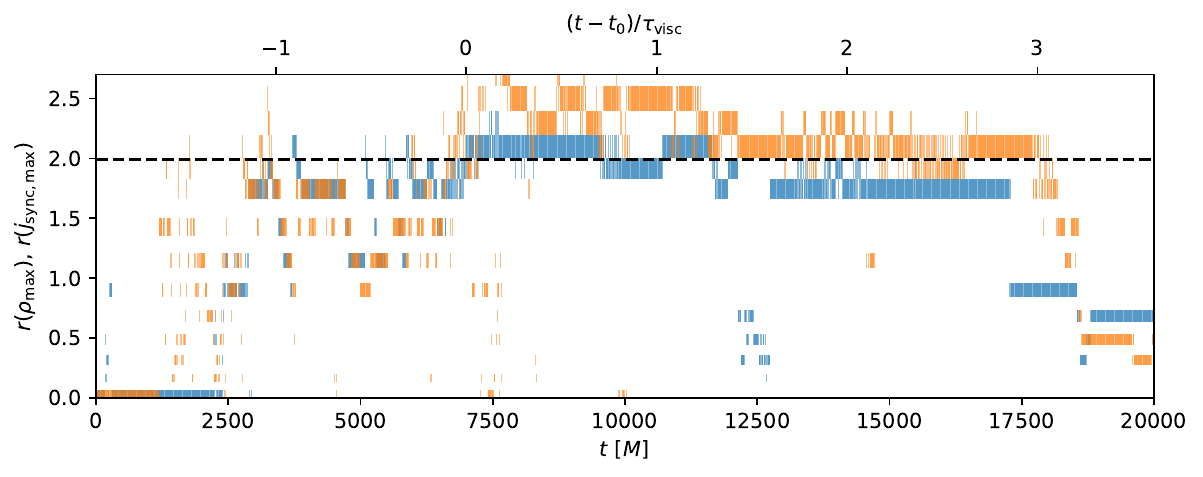}
    \includegraphics[width=0.9\linewidth]{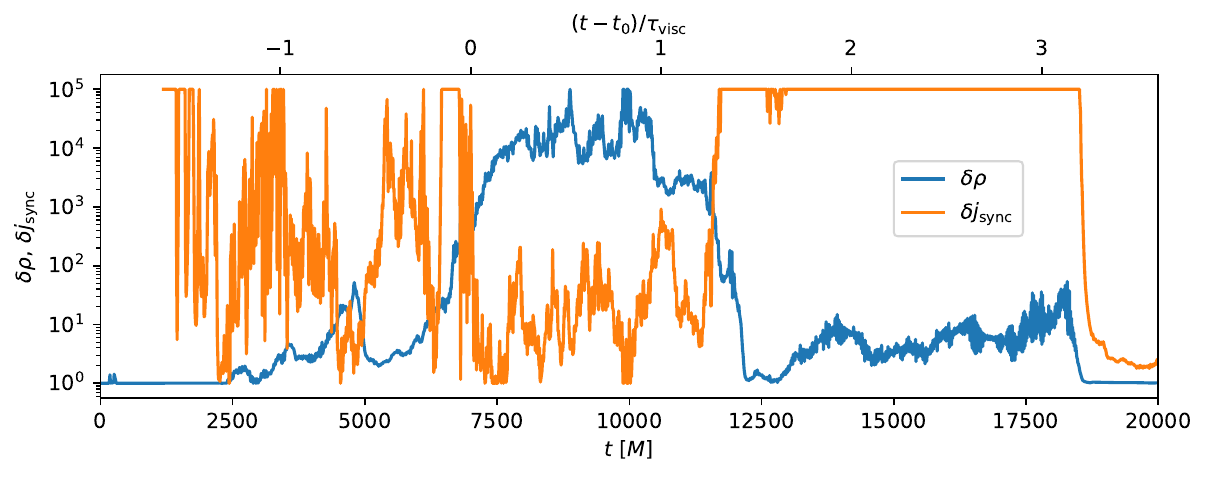}
	\caption{Top panel: evolution of the location of the local maxima of density $\rho$ and synchrotron luminosity proxy $j_{\rm sync}$ within $r<2.5\ M$. The prediction from the maximum $\Omega$ for circular geodesics ($r_{\rm max}=1.99\ M$) is shown as a dashed line. The vertical length of each rectangle is the radial extension of the computational cell containing the maximum.
    Bottom panel: evolution of the density and the luminosity contrast between the local maximum and the cell closest to the origin.
    In both panels, the upper horizontal axis shows time in units of the estimated numerical viscosity timescale (c.f. eq. \eqref{eq:viscous_timescale}).}
	\label{fig:rmax_vs_time}
\end{figure}

As a way to quantify the size of the hollow,
we keep track of the location of the density maximum within
the inner 2.5 $M$.
We calculate it by first obtaining the $\phi$-averaged profile of $\rho$ in the equatorial plane as\footnote{Due to the axial symmetry, we do not need to weight by the metric determinant.}

\begin{equation}
	\langle \rho \rangle (r) = \frac{1}{2\pi}\int_{0}^{2\pi} \rho(r,\theta=\pi/2,\phi)\ d\phi
\end{equation}
and then finding the radial coordinate of the local density maximum $\rho_{\rm max}$ within $r<2.5\ M$,
that we denote as $r(\rho_{\rm max})$.
The evolution of this quantity is shown in 
Fig.~\ref{fig:rmax_vs_time}
over the full duration of the simulation.
To reflect the discreteness of the numerical domain,
each data entry is represented by a rectangle
with a vertical size equivalent to the radial extension
of the computational cell that contains the density maximum, and a horizontal size of $10\ M$ (the output cadence).
To  quantify the
relative importance
of this maximum,
we calculate the ratios
$\delta\rho = \rho(r_{\rm max})/\rho(r_{\rm O})$
and 
$\delta j_{\rm sync}  = j_{\rm sync}(r_{\rm max})/j_{\rm sync}(r_{\rm O})$ 
of density and synchrotron 
emissivity $j_{\rm sync}$ 
between the location
of the maximum ($r_{\rm max}$ and the
cell closest to the
coordinate origin $r_{\rm O}$.
Synchrotron emissivity
is approximated
as
\begin{equation}
j_{\rm sync} \propto
\rho\ \Theta_{\rm e}^2\ \exp{\left[-0.2 \left(b\ \Theta_{\rm e}^2\right)^{-1/3}\right]}
\label{jsync}
\end{equation}
which is valid for an optically
thin plasma in the
frequency band accessible
to the EHT
\cite{leung_numerical_2011}.
Here $b$ is the
strength of the magnetic
field in the fluid frame,
and $\Theta_{\rm e} \coloneqq k_{\rm B} T/m_{\rm e} c^2$
is the dimensionless
temperature of electrons,
which we assume to be
the same as that of ions
in our simulation.
The ratio
$\delta j_{\rm sync}$
is invariant with respect
to changes in the density
scale of the simulation,
even if $j_{\rm sync}$ is not.

Before $t\sim 3\times10^3\ M$, the local density maximum
is located at the cell next to the origin.
During this time, the accretion flow has not reached
$r<25\ M$ and the local maxima reflect mainly the
atmosphere prescription and numerical noise.
Near $t\sim 3\times10^3\ M$, matter starts accumulating
close to the location expected from the $\Omega_{\rm max }$ prediction.
The local maximum then remains mostly within
$\pm 0.5\ M$ from the expected location until $t\sim 1.1\times10^4\ M$.
Deviations are indeed expected, as other mechanisms neglected
by the $\Omega_{\rm max }$ model are in play, such as pressure gradients,
turbulent velocity fluctuations and numerical viscosity.
Near $t\sim 1.1\times10^4\ M$, the density maximum starts
shifting slowly towards the origin.
Between $t\sim 1.45\times10^4\ M$ and $t\sim 1.65\times10^4\ M$,
it is possible to notice a competition between a local maximum near
$r=1.75\ M$ formed by incoming matter and another one near
$r=0.75\ M$ formed by matter that has been accumulating around the origin,
until the inner maximum dominates ($t\sim 1.65 - 1.85\times10^4\ M$),
to then shift towards the origin
near $t\sim 1.85\times10^4\ M$, and remain there until
the end of the simulation.

The evolution after $t\sim 1.1\times10^4\ M$ can be understood from the fact that in addition to
the MRI there are other mechanisms able to transport angular momentum,
although with longer associated timescales.

\subsubsection{Numerical viscosity and lifetime of the central hollow}
The fact that the plasma eventually
overcomes the centrifugal barrier and fills
the hollow necessarily means that there are
other processes in the simulation
that transport angular momentum,
although at a slower rate than the MRI.
The longer viscous timescale associated with these processes prevented the filling of the hollow from being observed in
\cite{Olivares:2018abq}, where the simulation was run for a shorter time.
(For comparison, the simulation for boson star model A reported there lasted until 10000~$M$, while in our simulation we observe matter penetrating the effective barrier around 17000~$M$, c.f. Fig.~\ref{fig:rmax_vs_time}).
Due to the length of the simulation,
it becomes relevant to estimate the
timescale associated to numerical
viscosity, which can produce the
necessary angular momentum transport
without any mechanism other than the MRI
being present in the simulation.
This does not exclude that such other processes are involved in the real system (e.g. molecular viscosity), although acting on much longer timescales.
%Since the relativistic corrections for distances and velocities
%are small and we 
The viscous timescale $\tau$
is an estimation of the
time required by viscosity $\nu$ to
diffuse mass or momentum over a characteristic
length $L$, and is given by

\begin{equation}
    \label{eq:viscous_timescale}
    \tau_\nu = \frac{L^2}{\nu} = \frac{L^2}{\ell_{\rm turb}}\frac{{\rm Re}}{u_{\rm turb}}\,,
\end{equation}
where in the last equality viscosity has been
rewritten in terms of the Reynolds number ${\rm Re}$, the typical  velocity associated to turbulence $u_{\rm turb}$ and the length scale at which energy is injected into the turbulent cascade $\ell_{\rm turb}$.

It has been shown \cite{shivakumar_numerical_2025} that Reynolds number associated to numerical viscosity can be estimated as

\begin{equation}
\label{eq:reynolds_no_numerical}
    {\rm Re} = \left[\frac{2\ell_{\rm turb}}{N_{\rm Re}}\frac{1}{\Delta x_p}\right]^{p_{\rm Re}} \,,
\end{equation}
where $\Delta x_p$ is the grid resolution,
$N_{\rm Re}$ is a constant given by simulations
such that if $\ell_{\rm turb}$ is resolved
by $N_{\rm Re}$ cells, ${\rm Re}=1$.
From numerical experiments, it has been found that $N_{\rm Re} \approx 0.7$ -- $1.5$,
and that $p_{\rm Re}$ is close to the theoretical value 4/3 for subsonic turbulence \cite{shivakumar_numerical_2025}.
We may perform an order-of-magnitude estimation of
the timescale associated to numerical viscosity
at $r=2\ M$
by considering both $\ell_{\rm turb} = L = 2\ M$.
Although the grid is not uniform due to the logarithmic
radial coordinate and the different AMR levels,
cells have approximately a square aspect ratio with
$\Delta x_p\approx 0.175\ M$ at that location.
For the typical velocity scale $u_{\rm turb}$, we choose the relative
velocity between circular geodesics at $r=2{\color{red}\ M}$ and
the value needed to co-rotate with circular geodesics near the origin, that is, $u_{\rm turb} = r\Delta\Omega_{\rm K} = 0.018$,
where $\Delta \Omega = \Omega_{\rm K}(r=2) - \Omega_{\rm K}
(r\rightarrow 0) $. In fact, before the dynamics is dominated by 
diffusion, the rotation profile in the disk is well approximated by 
circular geodesics. %\ho{Add figure.}
Substituting these values into equations
\eqref{eq:reynolds_no_numerical} and \eqref{eq:viscous_timescale},
we obtain ${\rm Re} = 65$ and $\tau_\nu=3604\ M$,
which is relevant for the duration of our simulation.

The upper horizontal
axis in Fig.~\ref{fig:rmax_vs_time}
shows time measured
in viscous timescales
from the time where
the density contrast
between the torus of
stalled matter and
the origin reaches $10^3$.
It can be seen that this
high density contrast
survives nearly one viscous
timescale, while the
location of the maximum
at the predicted
location survives for a
longer time.
This is an indication
that numerical viscosity
may be indeed the
process behind the filling
of the central hollow,
and that this viscous
diffusion would occur
on a much longer timescale
in a real system
in which the process
is associated to physical dissipation.

To get a sense of these timescales,
we can estimate that
corresponding to molecular
viscosity.
Although it is well known that
this type of viscosity results
in timescales of the
order of the age of the Universe
when applied to explain
accretion in systems
extending over hundreds of
astronomical units, it is
conceivable that it may become
important in diffusing matter
and momentum across the last
gap from $r_{\rm turn}$ to the
origin after the MRI has
driven accretion until that radius.
Molecular viscosity
can be approximated
as $\nu_{\rm m} = \alpha \lambda c_s$,
where
$\lambda$ is the mean free path
of molecules within the fluid,
$c_s$ is the sound speed,
and $\alpha$ is a factor of
order unity that depends on the
composition of the fluid, and
can be obtained in some cases
from statistical mechanics calculations.
After substituting this expression
in eq. \eqref{eq:viscous_timescale},
and setting again $L = r_{\rm turn}$,
we obtain
\begin{equation}
\label{eq:timescale-molecular-viscosity}
    \tau_{\nu} =
    1.5 \times 10^4\ {\rm years}
    \left( \frac{1}{\alpha} \right)
    \left( \frac{n}{10^{12}{\rm\ cm^{-3}}} \right)
    \left( \frac{\sigma_{\rm m}}{10^{-15} {\rm\ cm^{-3}}} \right)
    \left( \frac{0.1c}{c_s} \right)
    \left( \frac{r_{\rm turn}}{2r_g} \right)^2
    \left( \frac{M}{4\times10^6 M_\odot} \right)^2 \,,
\end{equation}
where $n$ is the particle
number density and
$\sigma_{\rm m}$ is the
cross-section of molecular collisions.
The numeric values
shown in the factors in parentheses in 
eq.~\eqref{eq:timescale-molecular-viscosity}
are reasonable
choices, albeit with a
high degree of uncertainty.
For $\sigma_{\rm m}$, we give an order-of-magnitude estimation
for proton-proton collisions
in the plasma.
For $n$, we choose a value
much larger than those
employed in one-zone models of
\sgrA ($\simeq 10^6\ {\rm cm^{-3}}$ \cite{Collaboration2022}) from
the consideration that the region around $r_{\rm turn}$ is accumulating matter over time. 
A more accurate estimation
is beyond the scope of this work,
however, it
should take into account the long-term evolution of mass accretion rate, outflows through barionic winds,
as well as viscous transport itself.
It is interesting to notice that,
although for an object of the mass of \sgrA the timescale to
diffuse matter to the center
is short in astrophysical terms,
it is much longer than
the time it takes to matter
to fall to $r_{\rm turn}$
over a comparable distance
from the region in which the MRI
is active ($\simeq$ a few minutes), which emphasizes
the fact that $r_{\rm turn}$
will be a region of accumulation.
Furthermore, the dependence
of eq. \eqref{eq:timescale-molecular-viscosity} on $n$
shows that as density increases
in that region, the timescale to
diffuse matter to the origin
will also increase.
It is also interesting to notice 
the quadratic scaling
of $\tau_{\nu}$ with the mass
of the object, which can increase
independently from the rest of
physical parameters in 
eq.~\eqref{eq:timescale-molecular-viscosity}.
This results in a timescale
of the order of $10^{8}$~years
for M87$^{*}$.

Finally, we briefly consider
other ways in which matter could
lose angular momentum 
matter to fall inside $r_{\rm turn}$.
Angular momentum could be
transported outward through spiral
shocks in the disks;
however, these are absent
in our simulation.
Another possibility is that
matter decelerates due to
magnetic fields lines
anchored in the disk,
which
transfer angular momentum to the
wind as in the process
described by Blandford and Payne
\cite{blandford_hydromagnetic_1982}.
This classic model
corresponds to a stationary,
self-similar Newtonian solution for a Keplerian,
geometrically thin accretion disk,
which limits its applicability
to our case of interest.
However, we may gain some qualitative
insight from the dependence
of the angular momentum transported
by the wind on
the magnetic field strength
(Section 5 of \cite{blandford_hydromagnetic_1982}).
The wind in our simulations appears
weakly magnetized, with
$\beta = p_{\rm gas}/p_{\rm mag} \gtrsim 0.1$, while we may
expect that the density in the disk
will keep increasing.
As matter accumulates near $r_{\rm turn}$, it will become increasingly hard for magnetic torques to
decelerate the disk.
The detailed evolution of
angular momentum transport in the
mini-torus will depend on the
interplay of 
angular momentum flux carried by
matter coming from the exterior
regions in the disk, and
that escaping through the wind.
It is unclear whether this process
would lead to filling the
central hollow in an observationally
relevant timescale, or would simply
contribute to removing excess
angular momentum from matter that
reaches $r_{\rm turn}$. 
A detailed study of this process
would require proper analytic
modeling, or long-term simulations reducing the effect of
numerical viscosity.
This uncertainty, leads to the
interpretation of
eq.~\eqref{eq:timescale-molecular-viscosity}
as an upper limit.

\subsubsection{Observational considerations}
In order to evaluate the
possibilities of $Q$-stars
acting as
black hole mimickers,
it is necessary not only to
verify the stalling of matter
at the predicted radius, but also
the formation of a bright ring
compatible to the expected size
of that measured by the EHT
Collaboration.
The upper panel of
Fig.~\ref{fig:rmax_vs_time}
shows that the location of the
maximum of our proxy for
synchrotron emissivity
(equation~\ref{jsync}) coincides
with the predicted radius during
most of the simulation,
and that the contrast with respect
to the origin is generally high
(bottom panel of 
Figure~\ref{fig:rmax_vs_time}).
Although a detailed analysis
on how this radius translates to the
size of the bright ring in images
would require to perform general relativistic
radiative transfer (GRRT) simulations,
we can obtain some
insight from the difference
between $b(r_{\rm turn})$
and the size of the observed ring
for the simulations
reported in \cite{Olivares:2018abq}.
The simulation of boson star~A
in such work shows a similarly
remarkable agreement between
the inner edge of the accretion disk
and the predicted $r_{\rm turn}$,
which results in an impact parameter 
for grazing photons
of $b(r_{\rm turn}) = 2.88\ M$.
The bright ring seen in such simulations
has a radius of 3.62~$M$ (edge on)
and
3.24~$M$ (face on), projected
on the image plane.
The radius of the bright ring is therefore
12\% -- 25\% larger than the impact parameter of grazing photons.
We may use this information
to make a reasonable prediction
for our case of interest.
For the $Q$-star used in our simulation,
$b(r_{\rm turn}) = 4.4\ M$.
From the estimates of mass
and distance assumed by the
EHT Collaboration 
\cite{Collaboration2022}
for \sgrA, $1\ M$
in projected distance
corresponds to an
angular size of 5.03~$\mu$as
on the observer's sky.
Using this conversion factor,
the impact parameter of the grazing
photons for a $Q$-star with the parameters
of \sgrA corresponds to an angular size
of 22.1~$\mu$as.
Therefore, we may expect a 
bright ring of angular diameter
49 -- 55.4~$\mu$as,
which can be compared to the
$51.8\pm 2.3$~$\mu$as
measured by the EHT Collaboration
\cite{collaboration_first_2022}.
This heuristic analysis suggests
that the size of the bright ring of these 
BH mimickers may be compatible
with current observations

%%%%%%%%%%%%%%%%%%%%%%%%%%%%%%%%%%%%%%%%%%%%%%%%%
\section{Discussion and conclusions}
\label{sec_discussion}

In this work we investigated whether stable $Q$-stars can act as viable black hole mimickers through the formation of effective shadow-like features. Our analysis combined a study of geodesic structure in self-interacting boson star spacetimes with three-dimensional GRMHD simulations of magnetized accretion flows.

A key result is the predictive power of the angular-velocity profile of timelike circular geodesics, Eq.~(5), as a first diagnostic for the expected morphology of the accretion flow. In astrophysical environments, matter surrounding a compact object typically forms a disk whose motion is well approximated by circular trajectories. The radial behaviour of $\Omega(r)$ therefore provides a proxy for the rotational properties of the plasma and for the conditions under which the magnetorotational instability (MRI) operates. In particular, the presence of a local maximum of $\Omega$ at a nonzero radius implies an inner region where $d\Omega/dr>0$, suppressing the MRI and favouring the formation of a stalled torus. Since synchrotron emission is sourced by magnetized rotating plasma, this mechanism naturally leads to a central luminosity depression.

Guided by this idea, we explored the parameter space of solitonic boson stars and identified families of $Q$-star solutions for which the angular-velocity profile develops such a turning point while remaining within a stable branch of the background solutions. In contrast with mini-boson stars - where similar behaviour typically occurs only in unstable configurations - we found that suitable combinations of quartic and sextic self-interactions allow for stable, relativistic configurations with sizable values of $r_{\rm turn}$ and of the corresponding impact parameter $b(r_{\rm turn})$. These configurations are therefore promising candidates to mimic the appearance of a black hole shadow.

We then performed GRMHD simulations of accretion onto a representative stable $Q$-star for which MRI suppression was expected to be active. The simulations confirm the qualitative picture suggested by the geodesic analysis. Matter transported inward by MRI-driven turbulence accumulates near the predicted location of $r_{\rm turn}$, forming a long-lived toroidal structure with a pronounced density and emissivity contrast relative to the central region. This configuration produces a low-density, low-luminosity core consistent with an effective shadow, demonstrating that stable bosonic stars can participate in the ``imitation game'' without requiring ultracompactness or an ad hoc disk prescription.

At later times, the simulations reveal additional dynamical effects. Although the MRI is suppressed in the inner region, other mechanisms ---such as turbulent transport and {\it numerical} viscosity--- enable angular momentum redistribution on longer timescales, allowing matter to slowly diffuse toward the center. In our simulations, this process eventually fills the hollow region, suggesting that the effective shadow produced by the stalled torus might not be absolutely stable but instead is a long-lived feature. It is worth emphasizing that in realistic astrophysical systems the physical viscosity is far smaller than the numerical viscosity, so this quasi-stable state could persist for significantly longer timescales.

Interestingly, the simulations also hint at a secondary mechanism capable of producing central brightness depressions. Even when the inner region becomes partially filled with matter, the magnetic field configuration can retain a spiral morphology and remain displaced from the center, generating a ring-like emission pattern through synchrotron radiation. The detailed physical origin and robustness of this mechanism require further investigation, but it may provide an additional pathway for horizonless compact objects to produce shadow-like images.

Overall, our results provide a proof of principle that stable $Q$-stars can mimic key observational features commonly associated with black holes. The effective shadows obtained here are somewhat smaller than those of a Schwarzschild black hole of the same mass, but arise from a qualitatively different mechanism that does not rely on the presence of a light ring or extreme compactness. This strengthens the case for considering self-gravitating scalar configurations as viable alternatives in the interpretation of horizon-scale observations.

Several directions for future work remain open. A systematic exploration of the $(g,h)$ parameter space could identify configurations yielding larger impact parameters and more pronounced effective shadows. Extending the analysis to rotating $Q$-stars is particularly important, given the relevance of spin for astrophysical sources. Incorporating radiative transfer calculations and producing synthetic images will allow for a more direct comparison with EHT observations. Finally, a deeper understanding of angular momentum transport in horizonless compact objects ---both physical and numerical--- will be essential to assess the longevity and observational relevance of the central brightness depressions reported here.

These developments will help clarify the extent to which horizonless compact objects can reproduce black hole-like phenomenology and, conversely, which observables may ultimately discriminate between the two scenarios.

%%%%%%%%%%%%%%%%%%%%%%%%%%%
%%%   ACKNOWLEDGMENTS   %%%
%%%%%%%%%%%%%%%%%%%%%%%%%%%

\acknowledgments
% Victor
VJ acknowledges support from the National Key R\&D Program of China under grant No.~2022YFC2204603.
% Hector
HO is supported by the Individual CEEC program - 5th edition funded by
the Portuguese Foundation for Science and Technology (FCT).
% Shuang-Yong
SYZ acknowledges support from the National Natural Science Foundation of China under grant No. 12475074 and No. 12247103.
% Darío
DN acknowledge the sabbatical support given by the Programa de Apoyos para la Superaci\'on del Personal Acad\'emico de la Direcci\'on General de Asuntos del Personal Acad\'emico de la Universidad Nacional Aut\'onoma de M\'exico in the initial stages of the present work.
% Computing grant
Simulations were performed on Mare Nostrum 5 under the FCT Advanced Compiting Project 2024.07059.CPCA.A3
 (\url{https://doi.org/10.54499/2024.07059.CPCA.A3}).
%% Other grants %%
% CONACyT
This work was partially supported by the CONACyT Network Project No.~376127 ``Sombras, lentes y ondas gravitatorias generadas por objetos compactos astrof\'\i sicos'',
% CIDMA
and also by the Center for Research and Development in Mathematics and Applications (CIDMA)  (\url{https://ror.org/05pm2mw36}) under the Portuguese Foundation for Science and Technology (FCT, \url{https://ror.org/ 00snfqn58}), Grants UID/04106/2025 (\url{https://doi.org/10.54499/UID/ 04106/2025}) and UID/PRR/04106/2025 (\url{https://doi.org/10.54499/UID/PRR/ 04106/2025}),
% FCT 
and by the FCT projects 2022.04560.PTDC (with DOI \url{https://doi.org/10.54499/2022.04560.PTDC}),
% CERN
as well as 2024.05617.CERN (with DOI \url{https://doi.org/10. 54499/2024.05617.CERN}). 
% NewFunFiCo
This work has further been supported by the European Horizon Europe staff exchange (SE) programme HORIZON-MSCA-2021-SE-01 Grant No. NewFunFiCO-101086251.

\appendix
\section{Central brightness depression in the final state}
\begin{figure}[ht!]
	\centering
	\includegraphics[width=0.7\linewidth]{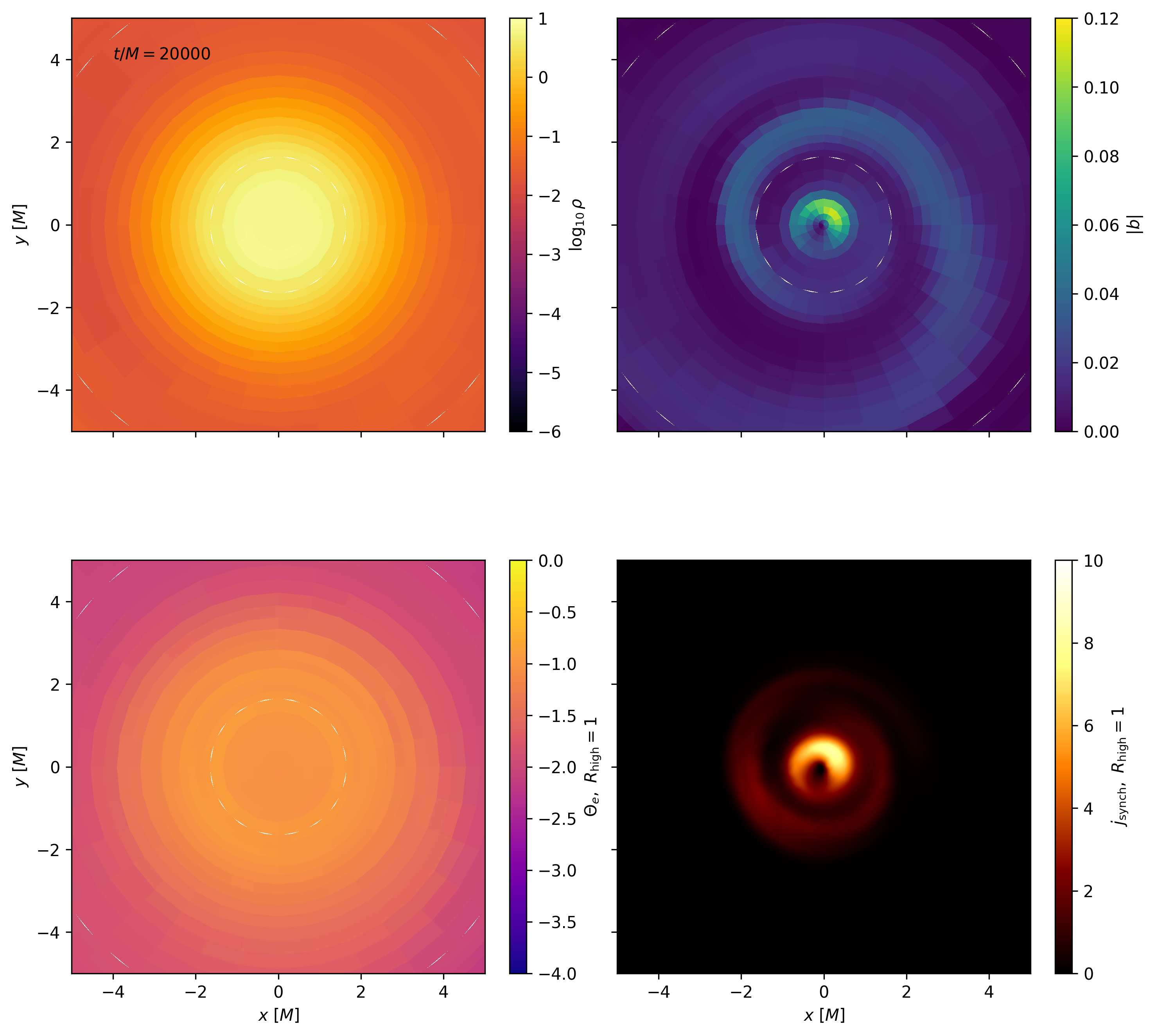}
	\caption{Snapshot of the simulation at $t=20000\ M$.}
	\label{fig:jproxy2d2000}
\end{figure}

In the final stages of the simulation, we observe that the central region becomes filled with matter. Although the timescale at which this happens is consistent with that of numerical viscosity associated to the discretization, it can be expected that this behavior will occur in the long run due to slower physical diffusion processes. For this reason, it is interesting to describe the final state of our simulation.
Fig.~\ref{fig:jproxy2d2000} shows this last snapshot for several quantities. It can be seen that density and pressure have reached nearly spherically-symmetric profiles. In contrast, the magnetic field has become `frozen' in a spiral configuration with the maximum displaced with respect to the center. As synchrotron emissivity is proportional to the strength of the magnetic field (c.f. eq.~ \eqref{jsync}), this configuration produces the ring-like emission pattern that can be seen in the botton right panel of the same figure.

At this stage, the MRI is inactive and Reynolds stress is negligible, as angular momentum has diffused and the rotation profile is nearly flat.
The magnetic field is small (magnetic pressure is much smaller than gas pressure) and effectively, it is advected passively, rotating together with the central structure. In the long run, numerical or physical resistivity are expected to diffuse it away, producing a more uniform configuration. Nevertheless, it is interesting to notice that this is a relatively long-lived state that produces a ring-like structure in luminosity, even though the central structure has a nearly spherically symmetric distribution in density.
This constitutes a possibility of producing ring-like images that, to our knowledge, has not been considered in the literature on exotic compact objects.
%%%%%%%%%%%%%%%%%%%%%%
%%%   REFERENCES   %%%
%%%%%%%%%%%%%%%%%%%%%%
%\bibliographystyle{bibtex/prsty}
\bibliography{ref}

%%%%%%%%%%%%%%%
%%%   END   %%%
%%%%%%%%%%%%%%%

\end{document}